\documentclass[12pt, a4paper]{article}
\usepackage{jcappub}

\usepackage[utf8]{inputenc}
\usepackage{amsmath, amsthm, amssymb, amsfonts}
\usepackage{bbold}
\pdfoutput=1
\usepackage{float}
\usepackage{booktabs}
\usepackage{graphicx}
\usepackage{subfig}
\usepackage{natbib}
\usepackage{hyperref} 
\usepackage{enumitem}

\usepackage[dvipsnames]{xcolor}
\usepackage{xcolor}
\usepackage{comment}
\usepackage{soul}
\usepackage{lineno}
\usepackage{url}

\usepackage{snapshot} %

\usepackage[normalem]{ulem}

\newcommand{\beq}{\begin{equation}}
\newcommand{\eeq}{\end{equation}}
\usepackage{bm}
\newcommand{\de}{{\textrm d}}

\graphicspath{{./figures/}}

\newcounter{notecountRJ}[section]

\def\lsim{\mathrel{\raise.3ex\hbox{$<$\kern-.75em\lower1ex\hbox{$\sim$}}}}
\def\gsim{\mathrel{\raise.3ex\hbox{$>$\kern-.75em\lower1ex\hbox{$\sim$}}}}

\newcommand{\rp}{\right)}
\newcommand{\lp}{\left(}

\def\bit{\begin{itemize}}
\def\eit{\end{itemize}}
\def\ben{\begin{enumerate}}
\def\een{\end{enumerate}}

\newcommand{\micron}{\mu \text{m}}

\newcommand{\be}{\begin{equation}}
\newcommand{\ee}{\end{equation}}

\newcounter{questioncount}[section]

\title{First constraints on QCD axion dark matter using James Webb Space Telescope observations}

\author[a,b,c,d]{Elena Pinetti}  
\emailAdd{epinetti@flatironinstitute.org}

\affiliation[a]{Center for Computational Astrophysics, Flatiron Institute, 162 5th Ave, New York, NY 10010}
\affiliation[b]{Emmy Noether Fellow, Perimeter Institute for Theoretical Physics,
31 Caroline Street N., Waterloo, Ontario, N2L 2Y5, Canada}
\affiliation[c]{Theoretical Physics Division, Fermi National Accelerator Laboratory,
Kirk and Pine St, Batavia, IL 60510
}
\affiliation[d]{Kavli Institute for Cosmological Physics, University of Chicago,
5640 S Ellis Ave, Chicago, IL 60637
}

\abstract{We present the first constraints on QCD axion dark matter using measurements from the James Webb Space Telescope. By utilizing publicly available MIRI and NIRSpec blank-sky observations, originally collected for sky subtraction purposes,  we derive strong limits on the axion-photon coupling constant $g_{a \gamma \gamma}$ in the mass range 0.1-4 eV. This analysis underscores the potential of blank-sky observations as a powerful tool for constraining dark matter models and demonstrates how astrophysical missions can be repurposed for particle physics research.}

\begin{document}

\maketitle

\section{Introduction} 
Unraveling the nature of dark matter (DM) is a fundamental challenge in modern physics. Although cosmological and astrophysical observations strongly support the existence of DM, its microscopic properties remain elusive (see Ref. \cite{Cirelli:2024ssz} for a review). Plausible beyond-the-Standard Model (BSM) theories suggest that DM could be unstable and decay into Standard Model particles on a cosmological timescale \cite{Slatyer:2017sev, Hooper:2018kfv, Safdi:2022xkm}. Among the BSM models proposing viable DM candidates, the axion stands out as one of the most promising options (see Ref. \cite{OHare:2024nmr} for a review). Initially introduced to resolve the strong charge-parity (CP) problem in quantum chromodynamics (QCD) ~\cite{1983PhLB..120..127P, 1983PhLB..120..133A, 1983PhLB..120..137D, PhysRevD.16.1791, PhysRevLett.38.1440, PhysRevLett.40.279, PhysRevD.11.3583}, these pseudoscalar particles can interact with Standard Model particles and decay into two photons. %
The theory of QCD axions has inspired various extensions of the Standard Model, proposing pseudoscalar DM candidates that possess similar couplings to photons but do not resolve the strong CP problem. These particles are referred to as axion-like particles (ALPs)~\cite{Adams:2022pbo}.
Due to energy conservation, eV-scale axions are expected to decay into infrared and optical photons. Space- and ground-based telescopes provide a promising means to detect such decay signals
(e.g. Refs.~\cite{Janish:2023kvi, Roy:2023omw}). Previous studies in this mass range have primarily focused on targeted observations of dwarf galaxies, globular clusters, and galaxy clusters \cite{Regis:2020fhw, Todarello:2023hdk, Grin:2006aw, Yin:2024lla, Caputo:2020msf, Carenza:2023qxh, Nakayama:2022jza, Ayala:2014pea, Dolan:2022kul,Kar_2025}.

The advent of the James Webb Space Telescope has provided a new avenue to constrain DM scenarios using blank-sky observations \cite{Janish:2023kvi}. 
These fields represent a significant untapped resource for DM studies due to the huge amount of data accessible. This is because blank-sky observations are routinely conducted for background subtraction in most astronomical surveys, meaning it is possible to use nearly the entire area surveyed by JWST instead of only the area around a few specific targets.  
Furthermore, the data are publicly available, allowing axion searches to be performed without requiring additional dedicated telescope time.

In this paper, we present the first constraints from JWST observations on axion dark matter down to masses of $0.1$ eV and rule out the QCD axion down to masses of $0.5$ eV.\footnote{At the masses of interest here, some QCD axion models are constrained by observations of supernovae \cite{Lella_2024,Springmann_2025,springmann2025universalboundqcdaxions} and neutron stars \cite{Buschmann_2022} but these bounds are evaded by flavor non-universal scenarios \cite{Di_Luzio_2018, Bj_rkeroth_2019, Badziak_2023}, permitting the existence of QCD axions with masses meV $\lesssim m_a \lesssim$ eV \cite{Co_2018, Caputo_2019, Visinelli_2010, Co_2020}.
}
This is the first time data from the JWST Mid-Infrared Instrument (MIRI) spectrograph have been used for this purpose. MIRI’s wavelength coverage of 4.9–28.8 $\micron$ enables the exploration of the QCD axion parameter space at lower masses than any previous ground- or space-based telescopes.
Besides using over 15524 MIRI measurements, we perform a multi-target analysis using over 1077 NIRSpec observations, which extends the mass range to 0.5-4 eV. Here, each observation refers to a single blank-sky exposure extracted from the JWST archive.
Since blank-sky observations are routinely acquired, the available dataset will increase steadily over the JWST mission lifetime, with an average of 4,800–7,200 for MIRI and 400–600 for NIRSpec annually.
Throughout this work, we adopt natural units where $\hbar = c = 1$, unless stated otherwise.

\section{Axion signal}
\label{sec:signal}
We searched for infrared emission lines originating from axion decays into two photons, within the Milky Way halo. 
The axion-photon interaction is described by a term in the Lagrangian 
\begin{equation}
\mathcal{L}_a = -\frac{1}{4} g_{a \gamma \gamma} a F_{\mu \nu} \tilde{F}^{\mu \nu} \; ,
\end{equation}
where 
$a$ denotes the axion field, $F_{\mu \nu}$ represents the electromagnetic field strength tensor and $\tilde{F}_{\mu \nu}$ is its dual. The strength of the axion-photon interaction is characterized by the axion-photon coupling g$_{a\gamma\gamma}$.
The expected observed flux for such an emission can be expressed as

\begin{equation}
\phi_{a} = 
  \frac{1}{4\pi} \frac{g^2_{a\gamma\gamma} m_a^3}{64 \pi} 
  \left( \frac{\de f}{\de \nu} * \mathcal{W} \right) \mathcal{D} \; ,
  \label{eq:axion}
\end{equation}
where $m_a$ is the axion mass.
The D-factor $\mathcal{D}\lp \theta \rp = \int\displaylimits \de s \, \rho $ is the DM column density for a given DM space density $\rho$, over the line-of-sight distance $s$.
The DM density profile is assumed to be a standard NFW profile with $r_s$ = 24 kpc and $\rho_s$ = 0.18 GeV/cm$^3$ \cite{Navarro:1995iw, Cirelli:2010xx, Cirelli:2020bpc}, and in Section \ref{sec:results} we discuss the uncertainty associated with a different choice of the DM profile. 
The emission spectrum $\de f/ \de \nu$  describes the fraction of total luminosity emitted in the frequency range $\de \nu$. The shape of the spectrum is largely determined by Doppler broadening and therefore depends on the DM velocity distribution. 
We model the DM velocity distribution as a uniform,
isotropic Maxwell–Boltzmann distribution with dispersion $\sigma_v$ = 160 km/s \cite{Evans:2018bqy}.  
The emission spectrum  $\de f/ \de \lambda$ is then a normal distribution with standard deviation $\sigma_0 = \lambda_0 \sigma_v$ and $\lambda_0 = 4 \pi / m_a$.
The observed flux is given by the convolution of the emission spectrum with the instrumental line spread function (LSF) $\mathcal{W}$, which we model as a Gaussian distribution in wavelength with width determined by the instrumental spectral resolution~\cite{gray2021observation, 2022A&A...661A..80J, NIRSpec-Dispersers-Filters, Janish:2023kvi}. The relation between the dispersion $\sigma_\lambda$ and the instrumental full width at half maximum $\Delta \lambda$ is given by $\sigma_\lambda = \Delta \lambda / 2 \sqrt{2 \, \textrm{ln}2}$. Hence, the observed line emission is modeled as a Gaussian distribution with variance $\sigma^2 = \sigma_\lambda^2 + \sigma^2_0$.
The flux is expressed in MJy/sr. As an illustrative example, in the case of MIRI, an axion with mass $m_a = 0.15$~eV and a coupling of $g_{a \gamma \gamma} = 5 \times 10^{-11}$~GeV$^{-1}$ observed toward $(b, \ell) = (9^\circ, 339^\circ$) produces a flux of 0.0016 MJy/sr. For NIRSpec, an axion with mass $m_a = 1$~eV and coupling $g_{a\gamma\gamma}=2\times10^{-10}$~GeV$^{-1}$ observed toward $(b,\ell)=(-21^\circ,295^\circ)$ produces a flux of $0.38$~MJy/sr.

\begin{figure}
    \centering
    \includegraphics[width=0.99\linewidth]{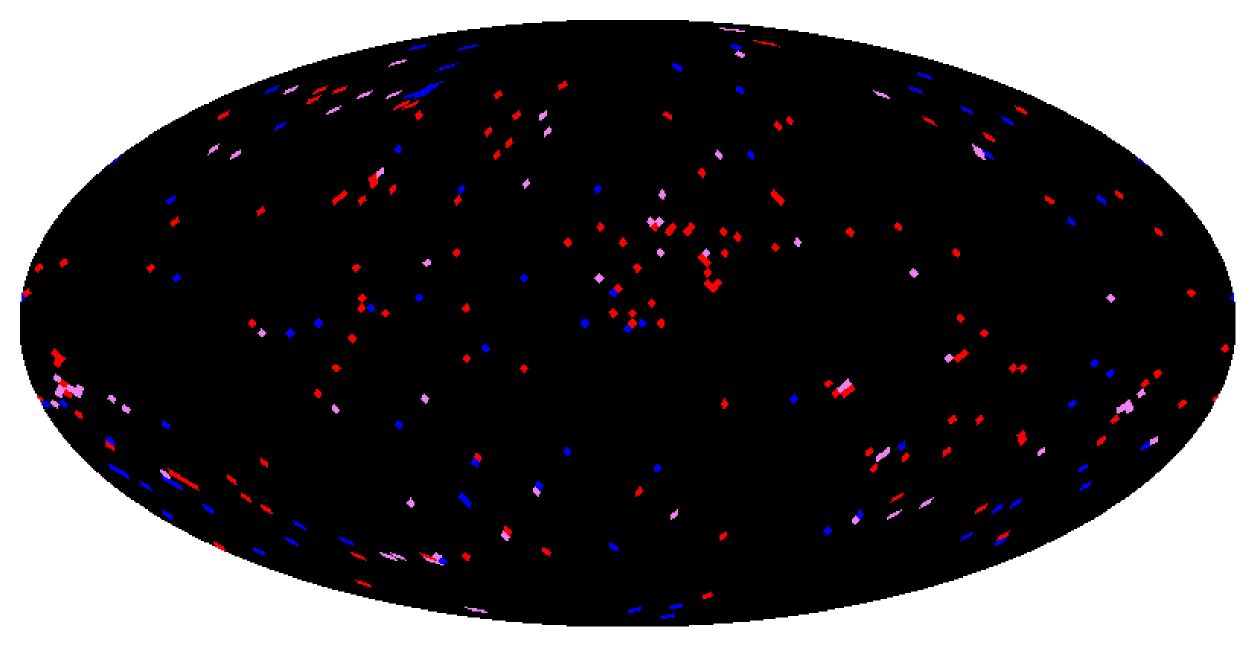}
    \caption{Map of the targets used in this analysis: targets observed both with NIRSpec and MIRI (purple dots), targets observed only with NIRSpec (blue), only with MIRI (red).}
    \label{fig:map}
\end{figure}

\section{James Webb Space Telescope Blank-Sky Observations}
\label{sec:JWST}

The James Webb Space Telescope (JWST) \cite{Gardner:2006ky} is an infrared observatory with instruments spanning the wavelength range (0.6-28.8 $\mu$m). 
In this analysis, we employ observations from two spectrographs on JWST, namely the Near-Infrared Spectrograph (NIRSpec, 0.6–5 $\micron$) and the Mid-Infrared Instrument (MIRI, 4.9–28.8 $\micron$). Both instruments provide integral field unit (IFU) spectrographs well-suited for blank-sky axion searches. NIRSpec covers 0.6 $\micron$ to 5.3 $\micron$ across a 3" × 3" field of view and includes high-resolution gratings ($ R \sim 2700$), three medium-resolution gratings ($R \sim 1000$), one low-resolution prism ($R \sim 100$) \cite{NIRSpec, gratings}. MIRI provides spectral resolution $R \sim 1500-3500$ over the wavelength range 4.9 to 28.8 $\mu$m and a 3.5" × 3.5" field of view \cite{2015PASP..127..584R, 2015PASP..127..646W, miri}.
The observations used in this analysis include 288 targets for NIRSpec, and 732 targets for MIRI. The sky coordinates of these observations are shown in Fig. \ref{fig:map}: the red dots refer to MIRI targets, the blue dots are NIRSpec targets, and the purple dots denote the targets that both instruments have observed.

In this analysis, we use Level 3 \texttt{science} products for both MIRI IFU and NIRSpec IFU observations, selecting only the \texttt{x1d} “Minimum Recommended Products”.\footnote{Minimum Recommended Products are the essential subset of JWST data products required for scientific analysis: \url{https://outerspace.stsci.edu/display/MASTDOCS/Minimum+Recommended+Products} \nocite{STScI:MRP}} These files contain the extracted 1D spectra. From them, we select the \texttt{BACKGROUND} and \texttt{BKGD\_ERROR} fields, which contain the flux-calibrated local background estimated by the JWST pipeline and its associated uncertainty. The \texttt{BACKGROUND} field is the same background that is subtracted to derive the source \texttt{FLUX} spectrum during the standard JWST calibration pipeline. For more details on the background components and the extraction procedure, we refer to the JWST documentation (see, e.g., \href{https://jwst-docs.stsci.edu/jwst-near-infrared-spectrograph/nirspec-observing-strategies/nirspec-background-recommended-strategies}{NIRSpec background strategies}; \href{https://jwst-docs.stsci.edu/jwst-mid-infrared-instrument/miri-operations/miri-dithering/miri-mrs-dedicated-sky-observations}{MIRI MRS dedicated sky observations}; \href{https://jwst-pipeline.readthedocs.io/en/stable/jwst/extract_1d/description.html}{pipeline extraction step \texttt{extract\_1d}}).

Table \ref{tab:nirspec} and Table \ref{tab:miri} report 10 example targets for NIRSpec and MIRI, respectively. Note that we utilized blank-sky observations taken as calibration sky flats for these target observations. For each blank-sky target in the table, the D-factor $\mathcal{D}$ and observed wavelength range are listed. All the MIRI and NIRSpec data used in this paper were downloaded from the MAST database \cite{mast}.
The specific observations used in this analysis are available through the MAST portal, at \href{https://archive.stsci.edu/doi/resolve/resolve.html?doi=10.17909/r3sx-9q78}{http://dx.doi.org/10.17909/r3sx-9q78} for MIRI and at \href{https://archive.stsci.edu/doi/resolve/resolve.html?doi=10.17909/m0s2-rh53}{http://dx.doi.org/10.17909/m0s2-rh53} for NIRSpec.

\begin{table*}
  \centering
  \begin{tabular}{@{}c c c c c c c@{}}
    \toprule
     Target name & \hspace{2.5mm} $ \lambda$ \hspace{2.5mm} & \hspace{2.5mm} D-factor \hspace{2.5mm}    \\
      & [$\mu$m] & [10$^{22}$ GeV/{cm}$^2$]  \\
    \midrule
    GC & 1.66-3.17 & 17.9 \\
    B335 & 2.87-5.27 & 2.35 \\
    IRAS 16253-2429 & 2.87-5.27 & 4.46 \\
    IRAS 20126+4104 & 2.87-5.27 & 1.45 \\ 
    HOPS370 & 2.87-5.27 & 0.876 \\
    LQAC 308-047 001 &  0.97-1.89 & 2.74 \\
    HE 0435-1223 & 0.97-1.89 & 0.913 \\ 
    HOPS153 & 2.87-5.27 & 0.882 \\
    RXCJ0600-z6 & 2.87-5.27  & 0.935 \\
    C/2014 UN271 & 0.60-5.30 & 1.30    
     \\
    \bottomrule
  \end{tabular}
  \caption{Examples of blank-sky NIRSpec observations. For each blank-sky target in the table, observed wavelength range (second column) and D-factor $\mathcal{D}$ (third column) are listed.}
  \label{tab:nirspec}
\end{table*}

\begin{table*}
  \centering
  \begin{tabular}{@{}c c c c c c c@{}}
    \toprule
     Target name & $\lambda$ & D-factor \\
     & [$\mu$m] & [10$^{22}$ GeV/{cm}$^2$]  \\
    \midrule
    IRAC J123711.88+622212.5 & 5.66-24.48 & 1.07 \\
    SDSSJ1652+1728 & 4.90-20.95 & 2.23 \\ 
    M-83-F1 & 4.90-28.6 & 2.04 \\
XID2028 & 4.90-20.95 & 1.04 \\ 
SPT0346-52 & 4.90-24.5 & 1.21 \\
C/2017 K2 & 4.90-28.6 & 3.58 \\
SPT-S J031134-5823.5 & 4.90-24.5 & 1.31 \\
IRAS 01364-1042 &  4.90-28.70 & 1.08 \\
SGAS J122651.3+215220 & 6.53-28.6 & 1.25  \\
2MASS J04202144+2813491 & 4.90-28.70 &  0.840   
     \\
    \bottomrule
  \end{tabular}
  \caption{Same as in Table \ref{tab:nirspec} but for MIRI.}
  \label{tab:miri}
\end{table*}

\section{Data analysis}
\label{sec:analysis} 

We search for an axion signal, setting bounds on the axion-photon coupling $g_{a \gamma \gamma}$ as a function of the axion mass $m_a$. The observed emission is modeled as the sum of two contributions: 
\begin{equation} 
\mathcal{E}(\lambda | g_{a \gamma \gamma}, \bm{\theta}) = \phi_{a}(\lambda, g_{a \gamma \gamma}) + \mathcal{B}(\lambda, \bm{\theta}) \; ,
\end{equation}
where $\phi_{a}$ represents an axion component, given by Eq. \eqref{eq:axion}, and
$\mathcal{B}$ is a background term, parameterized in terms of nuisance parameters $\bm{\theta}$. The blank-sky fields used in this analysis are off-target observations, therefore they are expected to be dominated by background emission. Indeed, the entire purpose of these blank-sky observations is to measure and subtract the local background emission from target observations. %
The goal of this work however is to find an axion line hidden in this overwhelming background. It is therefore crucial to remove as much astrophysical background as possible to maximize the sensitivity to axion lines.

The astrophysical background has four main components: zodiacal light,  detector thermal self-emission, stray light, and interstellar medium (ISM) background associated with the dust emission within the Milky Way. Unfortunately, there is not a sufficiently accurate parametric description of these backgrounds. The JWST backgrounds tool \cite{JWST-background} provides a rough estimate of these background emission components. However, a cursory comparison between the JWST tool predictions for these backgrounds and the actual measurements demonstrates that the tool does not provide a sufficiently accurate prediction to remove the backgrounds well enough to detect underlying axion emissions. 

 A more general approach to estimating the background is therefore necessary. Rather than assuming a parametric model for the background emission, we adopt a flexible approach, in which the spectrum of the total background is treated as a smooth polynomial \textit{locally} in wavelength, as described in detail in Appendix \ref{app:A}. The width of the region used for the polynomial fit is discussed in Appendix \ref{sec:sigma}. Since the reported
JWST uncertainties $\sigma_{i,j}$ underestimate the scatter observed in the data, we
calibrate them directly on the data products and account for residual
mismodelling when deriving the limits, as detailed in
Appendix~\ref{sec:error}. The robustness of the analysis is further tested in Appendix~\ref{sec:systematics}.
To prevent the analysis from being affected by possible astrophysical absorption lines, we exclude observations where an absorption line is detected, as explained in Appendix \ref{app:D}.

Following Refs. \cite{Cowan:2010js, Cirelli:2020bpc, Roach:2022lgo}, we define the goodness of fit $\chi^2$:

\begin{equation} 
     \chi^2 (g_{a \gamma \gamma}, \bm{\theta}) = \sum_{i,j} \left(\frac{\mathcal{E}(\lambda_{i,j} | g_{a \gamma \gamma}, \bm{\theta}_i) - d_{i,j}}{\sigma_{i,j}} \right)^2 \; ,
     \label{eq:chi2}
\end{equation}
where $\bm{\theta}$
are the nuisance parameters of the smooth background, and $d_i$ represents the JWST observations. The index $i$ runs over the dataset, while the index $j$ runs over the wavelength. 

\begin{figure*}
    \centering
    \includegraphics[width=0.99\linewidth]{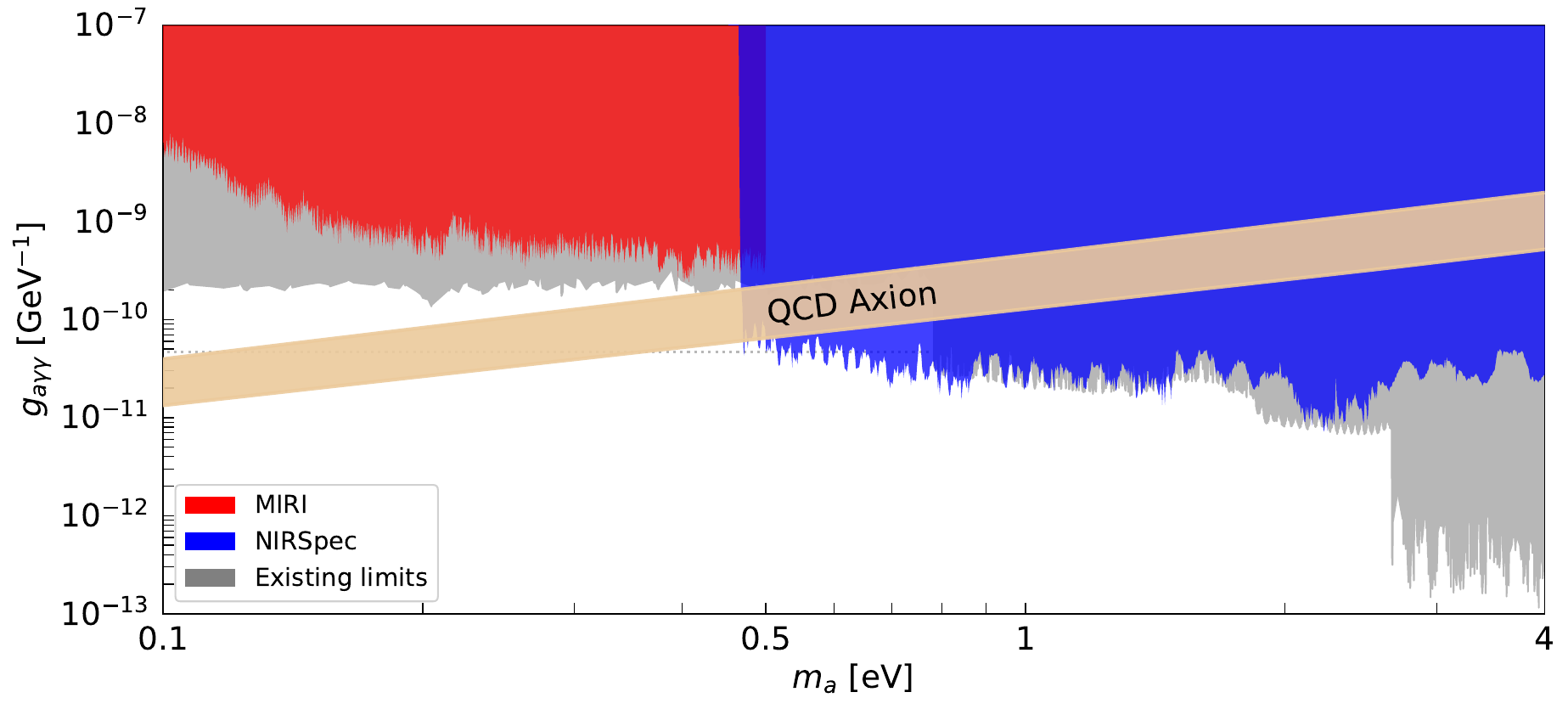}%
    \caption{New constraints on QCD axion dark matter and axion-like particles using MIRI (red) and NIRSpec (blue) blank-sky observations. The yellow band represents the parameter values where the QCD axion can comprise all the dark matter. Existing limits are discussed in the text.}
    \label{fig:gagg}
\end{figure*}

\section{Results}
\label{sec:results}
The 95\% constraints on the axion-photon coupling $g_{a \gamma \gamma}$ as a function of the axion mass $m_a$ are shown in Fig. \ref{fig:gagg}. These bounds assume that axions constitute the entire dark matter content of the Milky Way halo. The red line refers to the limits derived using 15524 blank-sky MIRI observations, while in blue we show the bounds obtained with 1077 sky flat measurements taken by NIRSpec. MIRI data constrain the mass range 0.1-0.5 eV, while NIRSpec limits higher masses between 0.5 and 4 eV.
The yellow band corresponds to the QCD axion band, i.e. the region of the parameter space where axions could simultaneously comprise the dark matter and solve the strong CP problem of quantum chromodynamics. 
The grey area is already constrained by indirect searches for axion decay with MUSE~\cite{Todarello:2023hdk},
WINERED \cite{Yin:2024lla}, 
and helioscope searches for axion-like particles by CAST~\cite{CAST:2007jps, CAST:2017uph}. 
Refs.~\cite{Ayala:2014pea, Dolan:2022kul} argue that the region above the dotted line is disfavored based on the impact of axion emission on stellar evolution. However, there is considerable complexity in setting such limits as shown in Ref.~\cite{2023ApJ...943...95S}.
Existing limits are drawn from Refs.~\cite{AxionLimits, Caputo:2021eaa}.

The results derived in this analysis have two important implications: 1) they represent the first constraints on axion dark matter using JWST observations and, more generally, infrared telescopes down to masses of 0.1 eV; 2) they improve over existing bounds for masses between 0.5 and 0.8 eV. At certain wavelengths, features consistent with lines are found with high significance. Many of these lines are consistent with known atomic and molecular lines. We leave to future work to explore their origin, and potentially mitigate their effect to allow for even stronger constraints to be derived.

\section{Conclusion}
\label{sec:conclusion}
This study constrains axion dark matter for the first time using measurements from the James Webb Space Telescope (JWST) down to masses of 0.1 eV, ruling out the QCD axion down to masses of 0.5 eV. We use publicly available blank-sky observations, originally gathered for sky subtraction purposes with the MIRI and NIRSpec spectrographs.
 
This blank-sky strategy enables future axion searches to be consistently updated and enhanced as JWST operations progress, since every observation will provide sky flats, regardless of the specific targets chosen for upcoming observations. This analysis demonstrates that blank-sky observations with infrared telescopes are a powerful tool for constraining dark matter models and for repurposing astrophysical missions for particle physics.

\acknowledgments
We thank Ryan Janish for help in obtaining the data used in this project. We thank Aurelio Amerio, Duncan Adams, Nikita Blinov, Andy Casey, Marco Cirelli, Neal Dalal, Chris Dessert, David Dunsky, Nicolao Fornengo, David Hogg, Junwu Huang, Konstantin Leyde, Gordan Krnjaic, Mariangela Lisanti, Karin Sandstrom, Oren Slone, Ken Van Tilburg and Francisco Villaescusa-Navarro for helpful conversations.
 
EP is grateful for the hospitality of Perimeter Institute where part of this work was carried out. Research at Perimeter Institute is supported in part by the Government of Canada through the Department of Innovation, Science and Economic Development and by the Province of Ontario through the Ministry of Colleges and Universities. 
The data presented in this paper were obtained from the Mikulski Archive for Space Telescopes (MAST) at the Space Telescope Science Institute (STSci). 
STScI is operated by the Association of Universities for Research in Astronomy, Inc., under NASA contract NAS5–26555. Support to MAST for these data is provided by the NASA Office of Space Science via grant NAG5–7584 and by other grants and contracts. We acknowledge support from {\sl Fermi Research Alliance, LLC} under Contract No. DE-AC02-07CH11359 with the U.S. Department of Energy, Office of High Energy Physics.\newline

\textbf{Note added.} Compared with v1, this revised version incorporates more recent JWST data products, which have updated and generally larger reported uncertainties. The exact list of products adopted is now available on the MAST portal.  This new version includes additional procedures to calibrate the reported uncertainties on the data products and accounts for the impact of residual mismodelling on the derived limits. Together, these changes yield more conservative constraints. During the final stages of the publication process, Ref.~\cite{Gemmell:2026xaa} appeared on arXiv, discussing related issues.

\bibliography{ref}%

\pagebreak

\appendix

\section{Polynomial Fit and Statistical Analysis}
\label{app:A}
The astrophysical
backgrounds include zodiacal light,
detector thermal self-emission, stray light, and interstellar medium (ISM) background associated with the Galactic dust emission.  
For each mass $m_{a}$, we describe the photon emission from axions as a Gaussian centered on wavelength $\lambda_a = 4 \pi / m_{a}$ with variance $\sigma^2 = \sigma_\lambda^2 + \sigma^2_0$, where $\sigma_\lambda^2$ is associated to the instrumental response and $\sigma^2_0$ includes the Doppler shift, as described in Section \ref{sec:signal}. The background is expected to be mostly smooth over the wavelength range covered by the axion line width. To avoid assuming a parametric form, we model the local background over a narrow range as a polynomial function:
\begin{equation}
p_n(\lambda) =  \sum^{n}_{k=0} c_k (\lambda - \lambda_a)^k  \; ,
\end{equation}
where the index $k$ runs from 0 to the polynomial order $n$, $c_k$ are the polynomial coefficients and $\lambda$ are the observed wavelengths in the spectrum. 
We fit this polynomial to the data which consists of a set of observed wavelength ${\lambda_i}$, their corresponding fluxes $d_i$ and error on the fluxes $\sigma_i$, with the index $i$ running over the data points within the narrow range centered on $\lambda_a$.
We can determine the best-fitting coefficients via the goodness of fit, $\chi^2$, defined as
\begin{equation}
    \chi^2 = \sum_{i} \left(\frac{p_n(\lambda_i) + \phi_a (\lambda_i) - d_i}{\sigma_i} \right)^2 \; ,
\end{equation}
where $\phi_a$ denotes the expected axion signal for a mass $m_a$ and axion-photon coupling $g_{a \gamma \gamma}$, as described in Section \ref{sec:signal}. For convenience, let us define $g_i = d_i - \phi_a (\lambda_i)$.
The best-fitting $c_k$ are the ones that minimize the $\chi^2$, and we can solve for them by setting $\partial \chi^2 / \partial c_k = 0 $ for all $k$. The resulting set of coefficients $c_k$ can be written compactly as 
\begin{equation}
    \vec{c} = \textbf{L}^{-1} \textbf{P} \, \vec{g} 
    \label{eq:cq}
    \; ,
\end{equation}
where $\vec{c}$ is the set of coefficients $\{c_k\}$ and $\vec{g}$ is the set of redefined fluxes  $\{g_i = d_i - \phi_a (\lambda_i)\}$. The  $(n+1) \times (n+1)$ matrix \textbf{L} has components given by 
\begin{equation}
    L_{kj} = \sum_i \frac{\left(\lambda_i - \lambda_a \right)^{k+j}}{\sigma_i^2} \; ,
\end{equation}
 and the $(n+1) \times h$ matrix \textbf{P} reads
 \begin{equation}
    P_{ki} =  \frac{\left(\lambda_i - \lambda_a \right)^{k}}{\sigma_i^2} \; ,
 \end{equation}
 where $h$ is the number of data points in the narrow range where the fit is performed, while $k$ and $j$ are indices running from 0 to $n$. 

Substituting Eq.~\eqref{eq:cq} back into the $\chi^2$ gives the profiled chi-square $\chi^2 = \vec{g}^{\,T} \mathbf{C}_m^{-1} \vec{g}$, where $\mathbf{C}_m^{-1} = \mathbf{C}^{-1} - \mathbf{P}^{T} \mathbf{L}^{-1} \mathbf{P}$ and $\mathbf{C} = \mathrm{diag}(\sigma_i^2)$. 
Here, $\mathbf{C}_m^{-1}$ is the projected inverse covariance obtained after profiling the polynomial in a Gaussian likelihood \cite{Tegmark:1997yq, Bridle:2001zv}.
Writing $\vec{\phi}_a = \Gamma\, \vec{a}$, the coefficients in Eq.~\eqref{eq:quad} are $A = \vec{a}^{\,T} \mathbf{C}_m^{-1} \vec{a}$ and $B = \vec{a}^{\,T} \mathbf{C}_m^{-1} \vec{d}$. Because $\mathbf{C}_m^{-1}$ annihilates any polynomial of degree $\leq n$, under the background-only hypothesis $B$ depends only on the noise and has mean zero. Since $\mathbf{C}_m^{-1} \mathbf{C} \, \mathbf{C}_m^{-1} = \mathbf{C}_m^{-1}$ \cite{Tegmark:1997yq}, its variance is $\operatorname{Var}(B)=A$. The best-fit amplitude is therefore $\hat\Gamma = B/A$ and has variance $1/A$, equal to the inverse Fisher information \cite{Cowan:2010js}. This result is exact here because the model is linear in both $\Gamma$ and the coefficients $c_k$. Consequently, under the assumption of Gaussian noise, the pull $z = B/\sqrt{A}$, i.e., the fitted amplitude in units of its uncertainty \cite{Demortier:2002pulls, ParticleDataGroup:2024cfk}, follows a standard normal distribution, provided that the data errors $\sigma_i$ correctly describe the true noise covariance and that the background is a polynomial of degree $\leq n$ within the fitting window. This is the result used in Section~\ref{sec:analysis}. Thus, if the measured distribution of $z$ has width $s\neq1$, the nominal amplitude uncertainties can be calibrated by rescaling them by a factor $s$.
 
 The order of the polynomial used in this analysis is $n=4$. 
 Appendix \ref{sec:sigma} discusses the choice of the width of the narrow region, while Appendix \ref{sec:systematics} discusses the impact of the choice of the order $n$ in the polynomial fit. 

\section{Width of the region in the polynomial fit}
\label{sec:sigma}
The polynomial function described in the previous section is computed locally, over a narrow range of wavelengths. The width of this range is taken as the coherence length of the background fluctuations. The main challenge is that the background contains line features similar in width to the axion signal, and it is crucial for this analysis that the background subtraction does not spuriously remove line emission. Therefore, it is necessary to remove these line features before determining the coherence length of the background.
The first step of this procedure requires identifying a line feature in the data to filter the signal out. This can be done by using something like a matched filter, that is by multiplying the data by the expected line shape
\begin{equation}
    \int \de \lambda \, \mathcal{F}(\lambda | \lambda_0) \, d(\lambda) \; ,
    \label{eq:int}
\end{equation}
where $d(\lambda)$ represents the data, the filter $\mathcal{F}$ is the shape of the axion flux, determined by $\left( \frac{\de f}{\de \nu} * \mathcal{W} \right)$ and $\lambda_0$ stands for the center of the line. We compensate the filter such that for a constant measured flux Eq. $\eqref{eq:int} = 0$. Therefore, we define the quantity
\begin{equation}
    Q(\lambda_0) = \int \de \lambda \, \mathcal{G}(\lambda | \lambda_0) \, d(\lambda) \; ,
    \label{eq:Q}
\end{equation}
where $\mathcal{G} = \mathcal{F} - \mathcal{K}$, and the function $\mathcal{K}$ is a 1D tophat function centered on $\lambda_0$ such that  $\int \de \lambda \, \mathcal{F}(\lambda | \lambda_0) = \int \de \lambda \, \mathcal{K} (\lambda | \lambda_0)$. Therefore, if the observed flux $d$ is constant, $Q=0$. This way the function $Q$ identifies lines and not the smooth background.
 The width of $\mathcal{K}$ is given by 10$\sigma$ ($\pm 5 \sigma$  centered on $\lambda_0$), where $\sigma$ is the root mean square of the assumed Gaussian profile of the DM line. Different choices for the width of $\mathcal{K}$ ($\pm 3 \sigma$ and $\pm4 \sigma$) lead to variations in the resulting bounds of about $3\%$ on average. 

The second step of this procedure requires the identification of a threshold above which we label a feature as a line to be removed. 
The variance of the filtered spectrum is
\begin{equation}
\sigma_Q^2(\lambda_0) = \langle Q^2(\lambda_0) \rangle - \langle Q(\lambda_0) \rangle^2 \; .
\label{eq>sigmaQ}
\end{equation}
Note that in the case of a pure background with no line features, $\langle Q \rangle = 0$. 

From Eq.\ \eqref{eq:Q}, it follows
\begin{equation}
    Q^2(\lambda_0) =  \int \de \lambda  \int \de \lambda' \; \mathcal{G}(\lambda|\lambda_0)  \, d(\lambda) \,  \mathcal{G}(\lambda'|\lambda_0)  \, d(\lambda')  \; ,
\end{equation}
thus

\begin{equation}
    \sigma_Q^2(\lambda_0)= \langle Q^2(\lambda_0) \rangle - \langle Q(\lambda_0) \rangle^2 \ =  \int \de \lambda  \int \de \lambda' \; \mathcal{G}(\lambda|\lambda_0)  \, \mathcal{G}(\lambda'|\lambda_0) \, {\rm Cov}(\lambda,\lambda') \, .
\end{equation}

We assume that the instrumental statistical measurement noise has diagonal covariance, hence ${\rm Cov}(\lambda,\lambda')$ is a diagonal matrix. We stress that this assumption applies only to the measurement noise and not to the astrophysical background. Strictly speaking, the measurement noise is not perfectly diagonal, since the JWST reduction pipeline introduces correlations between neighboring spectral bins. However, the standard Level 3 (`x1d`) spectral products provide per-wavelength variances but not the full covariance matrix, so we adopt the diagonal approximation, as is commonly done in analyses based on archival spectra \cite{Regis:2020fhw,Cirelli:2020bpc,Roach:2022lgo}.
Because the spectrum is observed at discrete wavelengths, we rewrite the integral as a sum:

\begin{align}
    \sigma_Q^2 &\approx \sum_{i,j} \mathcal{G}(\lambda_i|\lambda_0) \, \mathcal{G}(\lambda_j'|\lambda_0) \, {\rm Cov}(\lambda,\lambda') \Delta \lambda_i \, \Delta \lambda_j \nonumber \\
    &= \sum_{i,j} \mathcal{G}(\lambda_i|\lambda_0) \, \mathcal{G}(\lambda_j'|\lambda_0) \, \sigma_i^2 \, \delta_{ij} \, \Delta \lambda_i \, \Delta \lambda_j \nonumber \\
    &= \sum_i \mathcal{G}^2(\lambda_i|\lambda_0) \, \sigma_i^2 \, \Delta \lambda_i^2 \; ,
    \label{eq:Q2}
\end{align}

where $\sigma_i$ is the error bar on the measured flux and $\Delta \lambda$ is the wavelength pixel width. 

We remove lines when $Q(\lambda_0)$ differs from zero by more than $\pm 2 \sigma_Q$ (to remove both absorption and emission lines) and replace the removed fluxes by linearly interpolating between the edges of the excised region. 
We define $\mathcal{D}$ as the new dataset with the lines removed.
The autocorrelation function of $\mathcal{D}$ is
\begin{equation}
    \xi(\Delta \lambda) = \langle \mathcal{D} (\lambda) \, \mathcal{D} (\lambda + \Delta \lambda) \rangle - \langle \mathcal{D} \rangle ^2 \; ,
\end{equation}
where $\langle ... \rangle$ indicates the average over all wavelengths and observations. The variance of the new dataset is $\sigma^2_D = \xi(0)$. We define the coherence scale as the separation at which the correlation function $\xi$ is half of the variance $\sigma^2_D$. We find an average coherence scale that varies from 0.17 $\sim$ 0.25 $\mu m$ (depending on $\lambda_0$) for NIRSpec observations and 0.32 $\sim$ 1.21 $\mu m$ for MIRI blank-sky fields. This is the scale we use for subtracting the local background when searching for an axion line and determining the limits on the axion-photon coupling $g_{a \gamma \gamma}$. We also tested the dependence of the results on the threshold used to identify line features. Varying the threshold from $\pm1\sigma_Q$ to $\pm3\sigma_Q$ changes the final constraints by less than $3\%$ on average, indicating that the derived limits are resilient to this choice.

\section{Data-driven error calibration}
\label{sec:error}
Since the flux of Eq.~\eqref{eq:axion} is linear in the decay rate, we write $\vec{\phi}_a=\Gamma\,\vec{a}$, where $\Gamma \propto g_{a\gamma\gamma}^2$ and $\vec{a}$ is a fixed template. The
background coefficients $\bm{\theta}$ also enter linearly and can therefore be
profiled analytically \cite{Tegmark:1997yq, Bridle:2001zv} (see
Appendix~\ref{app:A}). Consequently, the profiled $\chi^2$ for each observation is exactly quadratic in the signal strength:
\begin{equation}
    \chi^2_i(\Gamma) = \chi^2_i(0) - 2 B_i \Gamma + A_i \Gamma^2 \; .
    \label{eq:quad}
\end{equation}
The unconstrained best-fit amplitude is $\hat\Gamma_i = B_i/A_i$, with variance $1/A_i$. In the absence of a signal, $B_i$ is Gaussian-distributed with mean zero and variance $A_i$, so the pull $z_i = B_i/\sqrt{A_i}$ follows a standard normal distribution \cite{Cowan:2010js, ParticleDataGroup:2024cfk}. Observations are combined by
summing their quadratic coefficients, $A=\sum_i A_i$ and
$B=\sum_i B_i$. The 95\% confidence interval is then defined by $\Delta \chi^2 \leq 3.84$, relative to the physically constrained best fit $\hat{\Gamma}=\max(0,B/A)$ \cite{Cowan:2010js}.

The uncertainties of spectra extracted from JWST IFU cubes are known to be underestimated, because cube building introduces correlations between neighboring spaxels \cite{Law:2023drizzle}. Consistent with this, we find that the distribution of pulls evaluated at wavelengths away from $\lambda_a$ is broader than a standard
normal distribution. We therefore calibrate the noise model directly from the data in three steps. Each correction is applied only when it increases the inferred uncertainty and is constructed to preserve the best-fit amplitude.

First, within each fitting window, the reported uncertainties are inflated to match the local scatter of the residuals about the background fit. This is the scale-factor prescription of Ref.~\cite{ParticleDataGroup:2024cfk}, applied locally as in Ref.~\cite{Janish:2023kvi}. Second, for each observation, the fit is repeated at 40 control wavelengths displaced from $\lambda_a$ by integer multiples of the line width. The corresponding $(A_i, B_i)$ are then rescaled so that the robust variance, estimated using the median absolute deviation, of the control-region pulls is unity, following the control-region calibration used in line searches (see e.g., \cite{Fermi-LAT:2013thd, Fermi-LAT:2015kyq}). Third, residuals shared by many observations at the same wavelength do not average down in the joint likelihood. We account for them using the ``spurious signal'' nuisance parameter of Ref.~\cite{ATLAS:2014cnc, Foster:2021ngm}. Profiling over this nuisance parameter is equivalent to rescaling the joint coefficients $(A, B)$ by $1/(1 + A \sigma_\nu^2)$, where $\sigma_\nu$ is determined from the scatter of the joint pull over the neighboring masses in the scan.

Finally, to prevent downward background fluctuations from producing artificially strong constraints, we apply the power-constrained limit prescription of Ref.~\cite{Cowan:2011an}, as in Ref.~\cite{Janish:2023kvi}.

\section{Tests for Systematics}
\label{sec:systematics} 
In this section, we estimate the impact of various systematics related to the background and axion modeling.

\vspace{0.5em}
\textbf{Background uncertainty.} 
As described in Appendix \ref{app:A} and \ref{sec:sigma}, the background emission is described by a polynomial fitted over a region of width given by the coherence length. This analysis assumes a polynomial of order 4.
We assess the sensitivity of the result to the polynomial order by considering polynomials of order 3 and 5. The axion limits change by no more than $10\%$ on average.

\vspace{0.5em}

\textbf{Dark matter density profile uncertainty.} 
The constraints derived above depend on the assumed D-factor $\mathcal{D}$, which is essentially the projected surface density of dark matter (DM) in each sightline.  We estimated $\mathcal{D}$ using the Navarro-Frenk-White (NFW) profile with halo parameters $r_s = 24$ kpc, $\rho_s = 0.18 \, \textrm{GeV}/\textrm{cm}^3$ \cite{Navarro:1996gj}.
To assess the impact of uncertainty in the assumed halo DM profile, we repeat the analysis for two alternative DM profiles: the Einasto profile with parameters $r_s = 11$ kpc, $\rho_s = 0.43 \, \textrm{GeV}/\textrm{cm}^3$ and $\alpha=0.16$ \cite{1965TrAlm...5...87E}; and the Burkert profile with parameters $r_s = 12.67$ kpc and $\rho_s = 0.712 \, \textrm{GeV}/\textrm{cm}^3$ \cite{Burkert_1995}. 

The analysis shows that using the Burkert profile has a minor impact on the resulting constraints, with variations of less than $15\%$, while the Einasto profile strengthens the constraints by $35\%$ (NIRSpec) to $50\%$ (MIRI). 
In the mass range of 0.8–1.5 eV, where the Galactic Center's DM density plays a crucial role in setting the limits, the constraints can change by up to $40\%$ depending on the chosen profile. 

\vspace{0.5em}

\textbf{Sensitivity to individual targets.}
We use jackknife resampling to test whether the constraints are dominated by a small subset of the targets.  
We consider a few dark matter masses where the axion limits appear more oscillating (e.g. $m_{\rm DM} = 0.115$ eV and $m_{\rm DM} = 0.169$ eV for MIRI, $m_{\rm DM} = 1.034$ eV and $m_{\rm DM} = 1.113$ eV for NIRSpec). We compute the bounds by excluding one target at a time. We find that the distribution has a peak and a very narrow tail, which extends no further than 10\% of the mean value of the distribution. This suggests that the bounds are not dominated by a few targets. 

\vspace{0.5em}

\textbf{Astrophysical foregrounds and systematics in the galactic plane.}
The Galactic plane is expected to suffer from the most astrophysical contamination of the DM signal from foregrounds and other systematics like absorption, including both broadband absorption (e.g., from dust extinction) and absorption from atomic and molecular lines.  Dust extinction
decreases sharply at longer wavelengths, and for the lines of sight considered in the analysis, broadband dust extinction is unimportant in NIRSpec and MIRI \cite{Schlegel_1998, Green_2019}. The exceptions are near-infrared observations close to the Galactic Center \cite{Gao_2013}. 

Because dust extinction and other foreground systematics are expected to be most severe in the Galactic plane, we adopt a conservative, model-independent approach and repeat the analysis after masking all observations within $|b|<30^\circ$. For MIRI, the masking weakens the constraints by $\sim25\%$ on average and up to $\sim55\%$. For NIRSpec, the impact is more modest, with an average degradation of $\sim5\%$ and a maximum degradation of $\sim15\%$ in the $0.8$--$1.5$ eV range.

\section{Absorption lines}
\label{app:D} 
In the previous appendix, we considered the effects of broadband dust extinction.  Another type of absorption to consider is line absorption due to molecular and atomic lines, which can exist both inside and outside of the Galactic plane.
To prevent the axion bounds from being affected by possible astrophysical absorption lines, we exclude the observations
where an absorption line is present at $\lambda_0 = 4 \pi / m_{\rm DM}$ with a significance greater than 2$\sigma_Q$, where $\sigma_Q$ is given by Eq. \eqref{eq>sigmaQ}.
This procedure is conservative: a narrow absorption feature at the wavelength corresponding to a given axion mass can lead to spuriously strong constraints, whereas excluding the affected observation avoids this bias.
The first step requires the identification of an absorption line. We follow the same procedure of Appendix \ref{sec:sigma} up to Eq. \eqref{eq:Q2} to identify a line. In particular, the absorption lines correspond to emissions such that $Q(\lambda_0)$ differs from zero by more than $- 2 \sigma_Q$. For every $\lambda$, we search for an absorption line in a specific observation and, if present, we do not utilize that observation to compute the bound for that specific value of $\lambda_0$.

By contrast, we do not remove emission-like features, since they could correspond to the axion signal itself; removing them would therefore discard potential signal and artificially reduce the sensitivity to an axion emission line.
Using this criterion, the absorption-line removal weakens the bounds by about 5\% on average.

\end{document}